\documentclass[twocolumn,
showpacs,preprintnumbers,amsmath,amssymb]{revtex4}

\usepackage{graphicx}
\usepackage{amssymb,amsmath,amsthm}
\newtheorem{Thm}{Theorem}

\theoremstyle{definition}

\newcommand{\bra}[1]{{\left\langle #1 \right|}}
\newcommand{\ket}[1]{{\left| #1 \right\rangle}}

\newcommand{\T}{\mbox{$\mathrm{tr}$}}
\newcommand{\B}{\mbox{$\mathcal{B}$}}
\newcommand{\cH}{\mbox{$\mathcal{H}$}}
\newcommand{\cI}{\mbox{$\mathcal{I}$}}
\newcommand{\C}{\mbox{$\mathcal{C}$}}
\newcommand{\N}{\mbox{$\mathcal{N}$}}
\begin{document}

\title{Bound on remote preparation of entanglement from isotropic states}

\author{Soojoon Lee}\email{level@khu.ac.kr}
\affiliation{
 Department of Mathematics and Research Institute for Basic Sciences,
 Kyung Hee University, Seoul 130-701, Korea
}

\date{\today}

\begin{abstract}
Using the negativity as an entanglement measure,
we investigate the possible amount of remotely prepared entanglement.
For two identical isotropic states on two-qudit systems 12 and 34,
we calculate the average amount of entanglement remotely distributed on the system 13
by joint measurement on the system 24,
and show that
the remote preparation of entanglement by the generalized Bell-measurement is optimal
among rank-one measurements
if the isotropic states have a certain fidelity with a maximally entangled state
in higher dimensional quantum systems,
or if the fidelity of the isotropic states is greater than a certain value
depending on the dimension.
In addition, we construct
a measurement better than the generalized Bell-measurement
with respect to the remote preparation of entanglement
when the isotropic states have small fidelity.
\end{abstract}

\pacs{
03.67.Mn,  
03.67.Bg, 
03.65.Ud  
}
\maketitle

\section{Introduction}
Entanglement provides us with a novel correlation between two or more parties,
which cannot be explained by any classical theories.
In addition, the correlation can be successfully applied to
quantum communication,
which seems to be classically impossible.
Thus entanglement shared between several parties has been considered as
one of the most significant resources
in quantum information processing including
quantum teleportation~\cite{BBCJPW} and quantum key distribution~\cite{BB84,Ekert91,B92}.

In order to perform a faithful quantum communication procedure in a given quantum network,
first of all, it should be required to prepare sufficient entanglement shared between desired parties.
As a generalization of entanglement swapping~\cite{ZZHE,ZHWZ,BVK},
there exists a process to remotely distribute entanglement between different parties~\cite{GS,Gour,KL,LKS},
which is here called the {\em remote preparation of entanglement} (RPE).

\begin{figure}
\includegraphics[width=.9\linewidth]{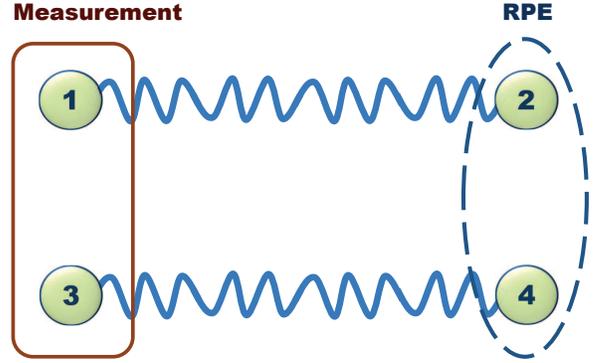}
\caption{\label{Fig:RPE}
A simple example of remote preparation of entanglement.}
\end{figure}

One of the simplest cases of the RPE is as follows (see FIG.~\ref{Fig:RPE}):
Assume that Alice (system 2), Bob (system 4) and the supplier Sapna (systems 1 and 3)
share an initial state $\rho^{12}\otimes\rho^{34}\in\B(\cH_1\otimes\cH_2\otimes\cH_3\otimes\cH_4)$,
and Sapna performs a joint measurement on systems 1 and 3.
Then entanglement can be probabilistically shared between Alice and Bob.

It has been known~\cite{GS,Gour,KL,LKS} that the possible average amount of the shared entanglement
is bounded by the product of entanglement amounts of $\rho^{12}$ and $\rho^{34}$.
In particular, if $\rho^{12}$ and $\rho^{34}$ are two-qubit states then
the bound can be shown in terms of the {\em concurrence} $\C$~\cite{Wootters} as follows~\cite{GS}:
\begin{equation}
\sum_{j=1}^s Q_j \C(\sigma^{24}_j)
\le\C(\rho^{12})\C(\rho^{34}),
\label{eq:GS}
\end{equation}
where $Q_j$ is the probability of the resulting state $\sigma^{24}_j$ on subsystems 24
after the joint measurement on subsystems 13,
and $\C(\rho^{12})$, $\C(\rho^{34})$ and $\C(\sigma^{24}_j)$ are
the concurrence of $\rho^{12}$, $\rho^{34}$ and $\sigma^{24}_j$, respectively.

However, we remark that
the maximal average amount of entanglement by the RPE
cannot generally attain to the bound,
the product of entanglement amounts of the two states.
Especially, for two-qubit states $\rho^{12}$ and $\rho^{34}$,
there does not in general exist a measurement to saturate the inequality~(\ref{eq:GS}).

We note that
if we can discover a measurement to achieve the maximal average amount of entanglement by the RPE
in a given quantum network
then we can learn how faithful quantum communication can be performed in the quantum network,
and this can be also employed
in constituting a quantum network for a desired level of quantum communication.
On this account,
it may be more important
to calculate the maximal average amount of entanglement obtainable by the RPE
or to find out the optimal measurement for the RPE
than to acquire a constant bound in a given quantum network.

In order to evaluate amount of entanglement,
we need an entanglement quantification.
Since the negativity is one of the most computable measures of entanglement
among several entanglement quantifications,
we here deal with the {\em negativity} $\N$~\cite{LKPL,VW,LCOK}, which is defined as
\begin{equation}
\N(\rho)\equiv \frac{\left\|\rho^{\Gamma}\right\|-1}{d-1}
\label{eq:negativity}
\end{equation}
for a two-qudit state $\rho$,
where $\left\|\cdot\right\|$ is the trace norm, and
$\Gamma$ is the partial transposition.
Moreover, since the negativity is an entanglement monotone~\cite{LKPL,VW,LCOK},
it can be regarded as a good measure of entanglement,
even though the bound entanglement with positive partial transposition~\cite{Horodecki1,Horodeckis2}
cannot be detected by the negativity.

In this paper, we investigate the possible amount of entanglement by the RPE
in terms of the negativity,
and show that, given two identical isotropic states as initial states,
the RPE by the generalized Bell-measurement is optimal
among rank-one measurements
if the isotropic states have a certain fidelity with a maximally entangled state
in higher dimensional quantum systems,
or if the fidelity of the isotropic states is greater than a certain value
depending on the dimension of quantum systems.
In addition, we show that
there exists a measurement better than the generalized Bell-measurement
with respect to the RPE
when the isotropic states have small fidelity.

\section{Remote preparation of entanglement and the generalized Bell-measurement}
In this section,
we show that the generalized Bell-measurement among rank-one measurements
is an optimal measurement
for the RPE presented in FIG.~\ref{Fig:RPE},
provided that two identical isotropic states with a certain fidelity are initial states.

Let
\begin{eqnarray}
\rho_F
&=& F\ket{\phi_d}\bra{\phi_d}
+ \frac{1-F}{d^2-1} \left(I\otimes I - \ket{\phi_d}\bra{\phi_d}\right)\nonumber\\
&=& F\Phi_d
+ \frac{1-F}{d^2-1} \left(\cI - \Phi_d\right)
\label{eq:isotropic}
\end{eqnarray}
be the isotropic state with fidelity $F$ in the two-qudit system,
where $\ket{\phi_d}=\sum_{j=0}^{d-1}\ket{jj}/\sqrt{d}$
is a maximally entangled state in the two-qudit system,
$I$ is the $d$-dimensional identity operator,
$\Phi_d=\ket{\phi_d}\bra{\phi_d}$ and $\cI=I\otimes I$.
We note that
the isotropic states form a generic class of two-qudit states,
to which any arbitrary two-qudit state can be transformed via
local quantum operations and classical communication,
and moreover it is not difficult to analyze their amounts of entanglement
computed by the negativity.
Hence quantum networks consisting of the isotropic states
are here dealt with.

We consider $\rho_{F}^{12}\otimes \rho_{F}^{34}$ as an initial state.
For the sake of convenience, we let $a=(1-F)/(d^2-1)$ and $b=(d^2F-1)/(d^2-1)$.
Then it is clear that $d^2a+b=1$, and
$\rho_{F}^{12}\otimes \rho_{F}^{34}$ becomes
\begin{eqnarray}
\varrho&\equiv&\rho_{F}^{12}\otimes \rho_{F}^{34}\nonumber\\
&=&
\left(a \cI^{12} + b\Phi_d^{12}\right)
\otimes
\left(a \cI^{34} + b\Phi_d^{34}\right).
\label{eq:rho_F0}
\end{eqnarray}

We now suppose that a rank-one measurement on subsystems 13 is performed,
and let $\ket{\psi}\bra{\psi}$ be a measurement operator of the rank-one measurement.
Due to the symmetry of isotropic states over local unitary operations,
without loss of generality, we may assume that
\begin{equation}
\ket{\psi}=\sum_{j=0}^{d-1}\sqrt{\lambda_j}\ket{jj}=\sum_{j=0}^{R-1}\sqrt{\lambda_j}\ket{jj},
\label{eq:general_psi}
\end{equation}
where $1\ge \lambda_0\ge\lambda_1\ge\cdots\ge\lambda_{R-1}> 0$,
and $\lambda_{R}=\lambda_{R+1}=\cdots=\lambda_{d-1}=0$.
It follows from straightforward calculations that
\begin{eqnarray}
 ^{13}\bra{\psi}\varrho\ket{\psi}^{13}
&=&
a^2 \cI^{24}
+\frac{b^2}{d^2}\sum_{i,j=0}^{R-1}\sqrt{\lambda_i\lambda_j}\ket{ii}^{24}\bra{jj}
\nonumber\\
&&+
\frac{ab}{d}\left(
\Lambda_R\otimes I
+I\otimes \Lambda_R\right)^{24},
\label{eq:rho_F2}
\end{eqnarray}
where
\begin{equation}
\Lambda_R=\sum_{j=0}^{R-1}\lambda_j\ket{j}\bra{j}.
\label{eq:Lambda_r}
\end{equation}
Then $\T \bra{\psi}\varrho\ket{\psi}=d^2a^2+b^2/d^2+2ab=1/d^2$.
Thus, the probability to obtain one measurement outcome after the rank-one measurement
is $1/d^2$, and
the resulting state $\rho_\psi$ in subsystems 24 after the measurement becomes
\begin{eqnarray}
\rho_\psi
&\equiv& d^2\cdot ^{13}\bra{\psi}\varrho\ket{\psi}^{13}\nonumber \\
&=&
d^2a^2 \cI^{24}
+b^2\sum_{i,j=0}^{R-1}\sqrt{\lambda_i\lambda_j}\ket{ii}^{24}\bra{jj}
\nonumber\\
&&+
dab\left(
\Lambda_R\otimes I
+I\otimes \Lambda_R\right)^{24}.
\label{eq:rho_F3}
\end{eqnarray}

We note that $A\otimes B\ket{\phi_d}=I\otimes BA^T\ket{\phi_d}$
for any operators $A$ and $B$ on the $d$-dimensional quantum system.
Hence, it can be clearly shown that
one outcome after performing an arbitrary rank-one measurement on subsystems 13
is always obtained with probability $1/d^2$,
and it can be also shown that
the resultant state in subsystems 24 is equivalent to the state $\rho_\psi$
up to local unitary operations,
if the pure state corresponding to the measurement outcome
has the same Schmidt coefficients as those of the state $\ket{\psi}$.

We now take into account the partial transpose on subsystem 4.
Then the partial transpose of $\rho_\psi$ is
\begin{eqnarray}
\rho_\psi^\Gamma
&=&
d^2a^2 \cI^{24}
+b^2\sum_{i,j=0}^{R-1}\sqrt{\lambda_i\lambda_j}\ket{ij}^{24}\bra{ji}
\nonumber\\
&&+
dab\left(
\Lambda_R\otimes I
+I\otimes \Lambda_R\right)^{24}.
\label{eq:rho_F4}
\end{eqnarray}
Thus, for each $0\le k\le d-1$, we have
\begin{equation}
\rho_\psi^\Gamma\ket{kk}=\left(d^2 a^2+(b^2+2dab)\lambda_k\right)\ket{kk}.
\label{eq:PT_eigenii}
\end{equation}
For each $0\le k<l\le d-1$, let
$\ket{\psi_{kl}^{\pm}}=(\ket{kl}\pm\ket{lk})/\sqrt{2}$,
then we also have
\begin{equation}
\rho_\psi^\Gamma\ket{\psi_{kl}^{\pm}}=
\left(d^2 a^2+dab(\lambda_k+\lambda_l)\pm
b^2\sqrt{\lambda_k\lambda_l}\right)\ket{\psi_{kl}^{\pm}}.
\label{eq:PT_eigenijp}
\end{equation}

It follows from Eqs.~(\ref{eq:PT_eigenii}) and (\ref{eq:PT_eigenijp}) that
$\ket{kk}$'s and $\ket{\psi_{kl}^{\pm}}$'s are eigenvectors of $\rho_\psi$,
and form an orthonormal basis for the two-qudit system.
Hence, we can obtain that the negativity of $\rho_\psi$, $\N(\rho_\psi)$, is
\begin{equation}
\frac{2}{d-1}\sum_{k<l< R}
\max\left\{0, b^2\sqrt{\lambda_k\lambda_l}-d^2 a^2-dab(\lambda_k+\lambda_l)\right\},
\label{eq:negativity_rho_psi}
\end{equation}
which is less than or equal to
\begin{equation}
\frac{2}{d-1}\sum_{k<l< R}
\max\left\{0, (b^2-2dab)\sqrt{\lambda_k\lambda_l}-d^2 a^2\right\}.
\label{eq:final}
\end{equation}
It is clear that Eq.~(\ref{eq:final}) becomes
\begin{equation}
\frac{2}{d-1}\sum_{(k,l)\in \mathcal{J}_\psi}
\left((b^2-2dab)\sqrt{\lambda_k\lambda_l}-d^2 a^2\right),
\label{eq:final2}
\end{equation}
where $\mathcal{J}_\psi$ is the set of all ordered pairs $(k,l)$
such that $k<l< R$ and $(b^2-2dab)\sqrt{\lambda_k\lambda_l}\ge d^2 a^2$.

Since $a=(1-F)/(d^2-1)$ and $b=(d^2F-1)/(d^2-1)$,
it can be shown that
if $d$ is sufficiently large to satisfy
\begin{equation}
b^2-2dab\ge 2(R-1) d^2 a^2
\label{eq:R_condition}
\end{equation}
then
\begin{eqnarray}
\N(\rho_\psi)
&\le&\frac{R-1}{d-1}\left(b^2-2dab-Rd^2 a^2\right)
\nonumber\\
&=&(b^2-2dab)\N\left(\Phi_R\right)-\frac{R(R-1)}{d-1}d^2 a^2
\nonumber\\
&=&\N(\rho_{\phi_R}).
\label{eq:final_ineq}
\end{eqnarray}
Furthermore, it can be also shown that
if the dimension $d$ is large enough to satisfy
\begin{equation}
b^2-2dab\ge (R+d-1)d^2 a^2
\label{eq:d_condition}
\end{equation}
then
\begin{equation}
\frac{R-1}{d-1}\left(b^2-2dab-Rd^2 a^2\right)
\label{eq:N(rho_{phi_R}}
\end{equation}
is less than or equal to
\begin{equation}
b^2-2dab-d^3 a^2,
\label{eq:N(rho_{phi_d}}
\end{equation}
that is,
the inequality $\N(\rho_{\phi_R})\le\N(\rho_{\phi_d})$ holds.
Thus, we obtain that
if the dimension $d$ is sufficiently large
such that
\begin{equation}
b^2-2dab\ge 2(d-1)d^2 a^2
\label{eq:d_final_condition}
\end{equation}
then
\begin{equation}
\N(\rho_\psi)\le \N(\rho_{\phi_d})
\label{eq:optimal_ineq}
\end{equation}
for any state $\ket{\psi}$.
Hence, it can be obtained that,
given a rank-one measurement of subsystem 13 with measurement operators $\{\ket{\psi_j}\bra{\psi_j}\}$
on the state $\varrho$,
if the inequality~(\ref{eq:d_final_condition}) holds then
\begin{equation}
\sum_j p(\psi_j) \N(\rho_{\psi_j})\le \sum_{s,t} p(\phi_{st})\N(\rho_{\phi_{st}}),
\label{eq:average_optimal}
\end{equation}
where $\rho_{\psi_j}$'s are the resultant states on subsystem 24 after the rank-one measurement,
$p(\zeta)$ is the probability to obtain $\zeta$ as a measurement outcome,
and $\ket{\phi_{st}}=I\otimes X^sZ^t\ket{\phi_d}$ with the generalized Pauli operators $X$ and $Z$
are the two-qudit generalized Bell states,
since $p(\psi_j)=p(\phi_{st})=1/d^2$ and $\N(\rho_{\psi_j})\le\N(\rho_{\phi_d})=\N(\rho_{\phi_{st}})$.

We remark that the left-hand side and the right-hand side in the inequality~(\ref{eq:average_optimal})
represent the average amount of entanglement of the resultant states after the rank-one measurement
and after the generalized Bell measurement, respectively.
This implies that, under the condition in the inequality~(\ref{eq:d_final_condition}),
the generalized Bell measurement is optimal among rank-one measurements
with respect to the RPE.

In addition, the inequality~(\ref{eq:average_optimal}) also holds
if $F$ is more than a certain value
depending on the dimension
so that
\begin{equation}
F\ge \frac{1+3d-d^2+(d^2-1)\sqrt{2d-1}}{d(d^2+2)},
\label{eq:F_condition}
\end{equation}
since the inequality~(\ref{eq:F_condition}) is equivalent to
the inequality~(\ref{eq:d_final_condition}).
Therefore, we can obtain the following theorem.
\begin{Thm}\label{Thm:main}
Assume that the initial states are two identical isotropic states
in the two-qudit quantum system.
Then the RPE by the generalized Bell-measurement
is optimal among rank-one measurements
if the dimension of the quantum system is large
enough to satisfy the inequality~(\ref{eq:F_condition}),
or if the isotropic states have fidelity more than a certain value
depending on the dimension
as seen in the inequality~(\ref{eq:F_condition}).
\end{Thm}

We remark that
for almost all initial states in higher dimensional quantum systems,
the RPE by the generalized Bell-measurement is optimal,
since the right-hand side in the inequality~(\ref{eq:F_condition}) tends to zero
as the dimension $d$ goes to the infinity.

\section{Measurements better than the generalized Bell-measurement}
In this section,
we show that
there exists a measurement to give higher entanglement than
the generalized Bell-measurement for the RPE
when the two identical isotropic states as initial states
have small fidelity.

\subsection{Three-dimensional quantum systems}
\begin{figure}
\includegraphics[angle=-90,width=.95\linewidth]{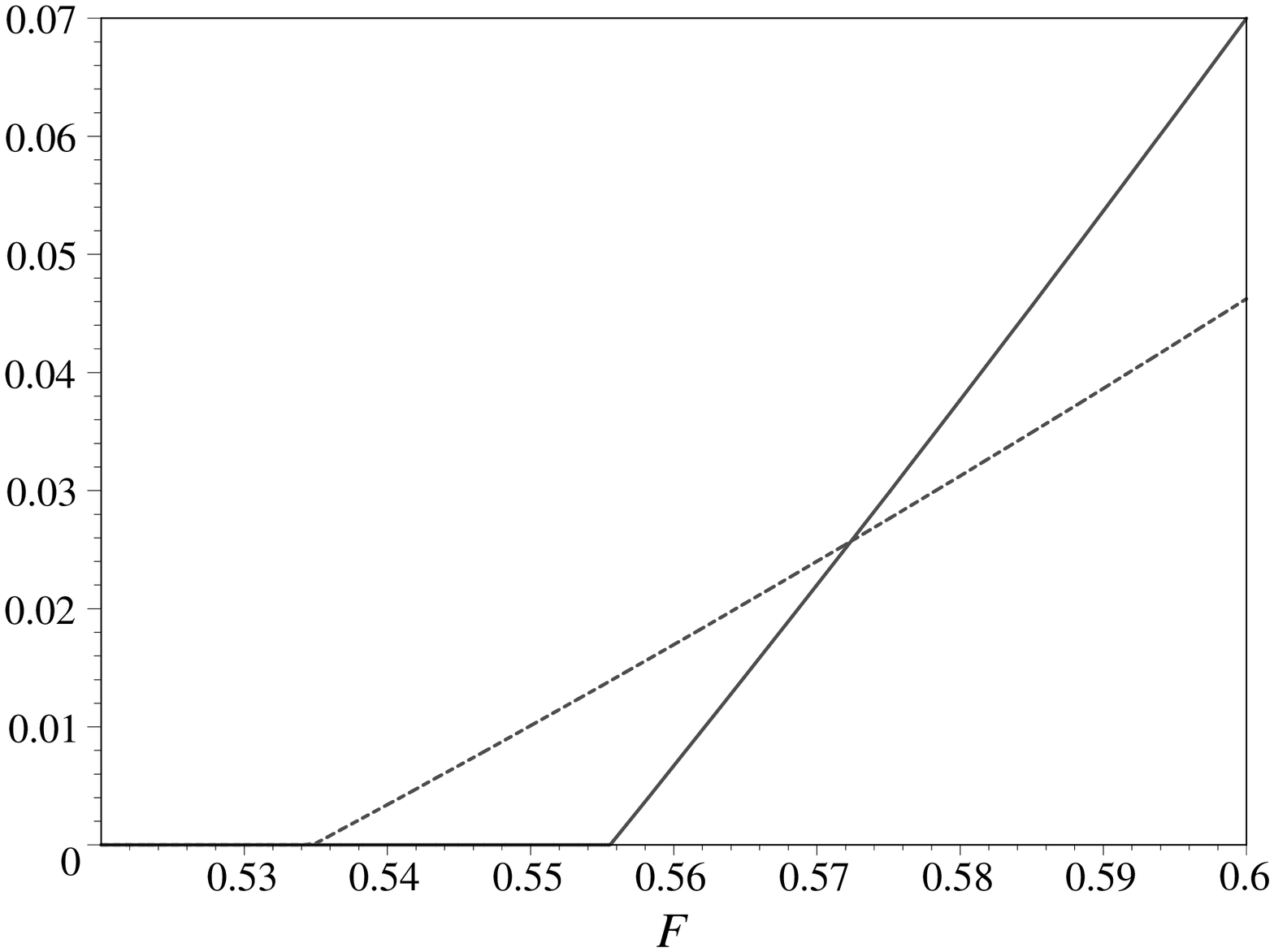}
\caption{\label{Fig:3d}
The values of $\N\left(\rho_{\phi_3}\right)$ and $\N\left(\rho_{\psi_{kl}^{\pm}}\right)$
in the qutrit systems:
The solid and dotted curves represent
the values of $\N\left(\rho_{\phi_3}\right)$
and $\N\left(\rho_{\psi_{kl}^{\pm}}\right)$, respectively:
$\N\left(\rho_{\psi_{kl}^{\pm}}\right)$ is more than $\N\left(\rho_{\phi_3}\right)$
for $0.535\approx\left(7+8\sqrt{3}\right)/39< F < \left(1+8\sqrt{5}\right)/33\approx 0.572$.}
\end{figure}

In this subsection, we assume that $d=3$.
Let $\omega_3=\exp(2\pi\mathrm{i}/3)$ with $\mathrm{i}=\sqrt{-1}$,
and consider a measurement whose measurement operators are
$\ket{\psi_{kl}^{\pm}}\bra{\psi_{kl}^{\pm}}$ for $0\le k<l \le 2$
and $\ket{\phi_{3}^{s}}\bra{\phi_{3}^{s}}$ for $0\le s\le 2$,
where
\begin{equation}
\ket{\phi_{3}^{s}}\equiv \sum_{j=0}^2 \omega_3^{sj}\ket{jj}.
\label{eq:phi_3k}
\end{equation}
Then it can be easily shown that
the measurement is a rank-one projective measurement,
and it can be also shown that
$\N\left(\rho_{\psi_{kl}^{\pm}}\right)$ is more than $\N\left(\rho_{\phi_3}\right)$
for
\begin{equation}
\frac{7+8\sqrt{3}}{39}< F < \frac{1+8\sqrt{5}}{33},
\label{eq:example_F_ineq_3d}
\end{equation}
as seen in FIG.~\ref{Fig:3d}.

It follows that
if the given isotropic states have fidelity $F$
satisfying the inequality~(\ref{eq:example_F_ineq_3d})
then the RPE by this measurement provides higher entanglement than
the RPE by the generalized Bell-measurement in terms of the negativity.

\subsection{$d$-dimensional quantum systems}
In this subsection, we generalize the example in the three-dimensional case.
Let $\omega_d=\exp(2\pi\mathrm{i}/d)$,
and consider a measurement whose measurement operators are
$\ket{\psi_{kl}^{\pm}}\bra{\psi_{kl}^{\pm}}$ for $0\le k<l \le d-1$
and $\ket{\phi_{d}^{s}}\bra{\phi_{d}^{s}}$ for $0\le s\le d-1$,
where
\begin{equation}
\ket{\phi_{d}^{s}}\equiv \sum_{j=0}^{d-1} \omega_d^{sj}\ket{jj}.
\label{eq:phi_dk}
\end{equation}
Then, as in the three-dimensional case,
the measurement is also a rank-one projective measurement.

It follows from tedious but straightforward calculations that
if the fidelity $F$ of the initial isotropic states satisfies
the following inequality
\begin{equation}
\frac{1-d+d^2+(d^2-1)\sqrt{3}}{d(d^2+2d-2)}
<F<
\frac{1+(d^2-1)\sqrt{d+2}}{d(d^2+d-1)},
\label{eq:example_F_ineq_general}
\end{equation}
then the measurement gives higher amount of entanglement than
the generalized Bell-measurement for the RPE.

From the two previous subsections,
we have the following theorem.
\begin{Thm}\label{Thm:example}
There exists a measurement such that
the RPE by the measurement has larger average amount of entanglement
than the RPE by the generalized Bell-measurement.
\end{Thm}

\section{Summary}
We have investigated the possible amount of entanglement by the RPE
in terms of the negativity,
and have shown that, given two identical isotropic states as initial states,
the RPE by the generalized Bell-measurement is optimal
among rank-one measurements
if the isotropic states have a certain fidelity with a maximally entangled state
in higher dimensional quantum systems,
or if the fidelity of the isotropic states is greater than a certain value
depending on the dimension of the quantum systems.

On the other hand, we have shown that
there exists a measurement better than the generalized Bell-measurement
with respect to the RPE
when the isotropic states have small fidelity.

\acknowledgments{
This work was supported
by a grant from the Kyung Hee University in 2011 (KHU-20110895).
}

\end{document}